\documentclass{acm_proc_article-sp}

\usepackage{algorithmic}
\usepackage{footnote}
\usepackage{url}
\usepackage{hyperref}
\usepackage{breakurl}
\usepackage[caption=false,font=footnotesize]{subfig}

\begin{document}
\title{Optimised hybrid parallelisation of a CFD code on Many Core architectures}

%
%
%
%
%

\numberofauthors{2} 
%
\author{
%
%
\alignauthor Adrian Jackson\\
\affaddr{EPCC} \\
\affaddr{The University of Edinburgh}\\
\affaddr{Mayfield Road}\\
\affaddr{Edinburgh}\\
\affaddr{EH9 3JZ, UK}\\
\email{Adrian.Jackson@ed.ac.uk}  
\alignauthor M. Sergio Campobasso \\
\affaddr{Department of Engineering} \\
\affaddr{Lancaster University} \\
\affaddr{Engineering Building} \\
\affaddr{Bailrigg} \\
\affaddr{Lancaster LA1 4YR, UK} \\
\email{m.s.campobasso@lancaster.ac.uk}
}
\maketitle
\begin{abstract}
Reliable aerodynamic and aeroelastic design of wind turbines, aircraft
wings and turbomachinery blades increasingly relies on the use of
high-fidelity Navier-Stokes Computational Fluid Dynamics codes 
to predict the strongly 
nonlinear periodic flows associated with structural vibrations and 
periodically varying farfield boundary conditions.
On a single computer core, the harmonic balance solution of the Navier-Stokes equations 
has been shown to significantly reduce the analysis runtime with respect
to the conventional time-domain approach. The problem size of realistic 
simulations, however, requires high-performance computing.
The Computational Fluid Dynamics COSA code features a novel harmonic balance
Navier-Stokes solver which has been previously parallelised using both 
a pure MPI implementation and a hybrid MPI/OpenMP implementation.
This paper presents the recently completed optimisation of both
parallelisations.
The achieved performance improvements of both parallelisations highlight
the effectiveness of the adopted parallel optimisation 
strategies.
Moreover, a comparative analysis of the optimal performance of
these two architectures in terms of runtime and power consumption 
using some of the current common HPC architectures
highlights the reduction of both aspects 
achievable by using the hybrid parallelisation with emerging many-core architectures. 
\end{abstract}


\section{Introduction}
\label{sec:0}

Reliable aerodynamic and aeroelastic design of wind turbine and rotorcraft 
blades, aircraft
wings and turbomachinery blades increasingly relies on the use of
high-fidelity Navier-Stokes (NS) Computational Fluid Dynamics (CFD) codes 
to predict the strongly 
nonlinear periodic flows associated with structural vibrations and 
periodically varying farfield boundary conditions.
Many examples of using NS codes for such simulations have been published 
\cite{ati_Sayma_Imregun_Simpson_2007,FLD:FLD1757}.
These simulations have high computational
costs even when parallel or high performance computing is adopted. This 
is particularly true for 
unsteady time-domain (TD) three-dimensional (3D) NS equations, 
which can require 
many months of computational time on a single computer.
The use of a frequency-domain (FD) technique can dramatically reduce the 
required runtime for simulation 
of unsteady periodic flows with respect to that of TD solvers.  
On a single computer core, the FD harmonic balance (HB) solution of the NS equations 
has been shown to reduce at least by an order of magnitude 
the analysis runtime with respect
to the conventional time-domain approach\cite{Campobasso_Baba-Ahmadi_2011,enlighten44055}.
The problem size of realistic periodic flow problems,  
however, is such that, despite the above discussed reduction,
the HB NS analysis still requires high-performance computing.

Computational hardware has been rapidly evolving over recent years, 
particularly with 
the advent of multi-core processors and many-core hardware.  Clusters of 
shared-memory servers have been common 
for high-end computational resources but the advent of multi- and many-
core processors 
is bringing significant shared-memory computational resources at the node 
level for these clusters.
Distributed memory parallelisations (based on the message passing library 
MPI~\cite{mpi2}) 
have been the primary method for parallelising scientific codes, such as 
the one considered in this paper,
for the past fifteen years; primarily due to the availability of large 
distributed memory systems and the prohibitive 
cost of large shared memory systems. Shared memory parallelisations, 
generally undertaken using the OpenMP~\cite{openmp} 
shared memory library, have been restricted to a number of specialised 
high performance computer (HPC) systems 
or to very small numbers of processors.  

The CFD structured multi-block 
COSA code features a novel HB NS 
solver which has been previously parallelised using both 
a pure MPI implementation and a hybrid MPI/OpenMP implementation\cite{DBLP:journals/ife/JacksonC11,on_Campobasso_Baba-Ahmadi_2011}.
This paper presents the recently completed optimisation of both
parallelisations aiming to optimise performance and thereby reduce time to solution, or 
improve computational efficiency, for a given simulation. 

An increasing number of research programmes 
aimed at developing efficient hybrid parallelisation technologies
are under way.
The OP2 library\cite{Mudalige_op2:an} provides users with functionality 
to implement CFD applications using 
unstructured meshes on a range of different computational hardware, 
including multi-core and many-core (particularly GPGPU) systems.  OP2 
undertakes the parallelisation and distribution of work and data for 
users, however it is not directly applicable to structured multi-block 
codes. Multi-block structured codes do 
not require mesh partitioning software to run a simulation; 
they rather require a user to create the mesh partitioning separately to the 
simulation. Beside the OP2 project, a number of large scale CFD packages 
have recently investigated 
hybrid parallelisations, adding OpenMP to the 
existing MPI parallel functionality, including 
OpenFOAM\cite{OpenFOAMTHESIS} where both a task based parallelism and a 
standard data parallel (parallelisation of loops using OpenMP) have been 
added and been shown to improve performance.  Fluidity\cite{fluidityref}, 
along with other Algebraic Multigrid\cite{AMGhybrid2011} solver based CFD 
packages, has added OpenMP to parallelise the intensive task of 
constructing the matrix of equations to be solved, often coupled with a 
hybrid solver library such as PETSc\cite{weiland:mixed-mode}.    There 
has also been a strong move towards porting CFD codes to GPGPU 
architectures with a number of commercial code, such as ANSYS's 
Fluent\cite{fluentgpgpu} application, being ported to a range of GPGPU 
hardware.

Mixed-mode, or hybrid, parallelisation of scientific simulation code have 
been undertaken across a wide range of scientific disciplines, from 
gyrokinetic codes for plasma 
simulations\cite{Madduri:2011:GTS:2063384.2063415}, to molecular dynamics 
packages\cite{cp2khybrid}, and finite element simulations of electrical 
systems\cite{6324486}.  In general, all these efforts have provided 
performance improvements over the standard MPI or OpenMP parallelisations 
previously implemented in these applications when using shared memory 
clusters, although careful optimisation work is required to ensure 
performance improvements are realised.

The work presented in this paper builds on previous 
parallelisation work of the COSA HB NS
solver~\cite{on_Campobasso_Baba-Ahmadi_2011,DBLP:journals/ife/JacksonC11}, 
and has been undertaken 
to 1) improve the efficiency of the MPI parallelisation through 
rationalisation of the messages sent between processes and optimisations 
of the MPI-I/O utilisations, and 2) improve the hybrid parallelisation 
performance by extending the amount of functionality covered by the 
OpenMP implementation, and re-engineering the OpenMP code to reduce 
overheads and improve performance.

The harmonic formulation of the HB NS equations and the 
main structural features of the COSA code are summarised
in the next two Sections. We then describe the optimisations 
undertaken to the MPI and hybrid parallelisations in the following two 
sections, and finally discuss the benchmarking undertaken on 
a range of the worlds largest HPC resources, and draw some conclusions, 
in the final sections of the paper.

The OpenMP parallelisation can work for all three solvers, with different parallelisations available over the blocks in the multi-block grids, over the harmonics 
for the HB solver, and over the grid points for those problems that use low numbers of blocks or harmonics (for instance a single block, TD, simulation).  However, 
the OpenMP code is limited to the size of shared-memory machine available.

The MPI parallelisation distributes the blocks of the multi-block grid over the available MPI processes to distribute the work of the simulation. 
Communication is required between the blocks where data on the edge of blocks (called cuts in COSA) needs to be communicated to neighbouring blocks (halo communications).  
The maximum number of processes that the MPI parallelisation can use is limited by the number of geometric partitions (grid blocks) in the simulation.

The hybrid parallelisation combines the MPI code with either the harmonic OpenMP parallelisation or the grid point OpenMP parallelisation, depending on the simulation being performed.

\section{Optimisations}
\label{sec:4}

Of the three different parallelisations of COSA the MPI version is the one currently used for production simulations.  In general the efficiency of the original MPI implementation is acceptable,  for example with  
test case 1 (described in Section \ref{sec:5}) experiencing around 50\% efficiency when run on 512 cores (compared to running on a single core on a Cray XE6), or around 70\% efficiency if 
considering the performance without writing any data to file at the end of the simulation\footnote{This is when using a small number of simulation iterations, 100.
  Normal simulations would run with thousands of iterations}.

The hybrid parallelisation was implemented to enable further reduction in time to solution for a given 
problem beyond what could be achieved by using the pure MPI code by enabling the usage of more computational 
resources (by scaling beyond the number of blocks in the simulation).  The MPI code is limited in the 
number of processing elements, or cores, it can use by the 
number of blocks in the simulation.  For instance, for test 
case 1 and 2 that we used to benchmark COSA for this paper the MPI code is limited to 512 and 2048 cores 
respectively.  The hybrid code can enable 
further resource to be used, provided the HPC machine that the code is being run on is a shared memory 
cluster (as is generally true of modern HPC machines), by underpopulating each node with 
MPI processes and enabling each MPI process to create a number of OpenMP threads to utilise the cores 
left free of MPI processes.  If a machine is composed of 32-core nodes, such as the Cray XE6 
described in Section \ref{sec:5}, then we can, for example, place 4 MPI processes on to each node, 
let them each create 8 OpenMP threads, and we can utilise 128 nodes for a simulation with 512 blocks 
as compared to 16 nodes for the MPI code.  

However, the existing hybrid code does not provide optimal performance, and therefore resource usage.  For test case 1 we can reduce the runtime of the simulation by four times when using 
eight times the amount of resources as the MPI code can utilise.  Therefore, both the MPI and hybrid parallelisations were optimised to improve performance and thereby reduce the time to solution 
required to solve a given simulation on a given amount of computational resources, as outlined in the following subsections.

\subsection{MPI Optimisations}
The initial focus of our optimisation work was on the  MPI communications in COSA.  The current code utilises non-blocking MPI communications, but for a large simulation there can be as many as 5,000 messages 
sent between pairs of communicating processes at each Runge-Kutta step to communicate {\it halo} or cut data to neighbouring processes.  This is because the original implementation sends small parts of the boundary data 
to neighbouring processes at a time, with an example of this shown in the following pseudo code:

\begin{verbatim}
do i = 0,boundary length
   if(myblock1 .and. myblock2) then
      do n = 0, 2*nharms
          do ipde = 1, npde
              copy 1st part of q2 to q1
	      copy 2nd part of q2 to q1
          end do
      end do
   else if(myblock1) then
      receive 1st part of q1 from remote process
      receive 2nd part of q1 from remote process
    else if(myblock2) then
      send 1st part of q2 to remote process
      send 2nd part of q2 to remote process
   end if
end do
\end{verbatim}

Note that in the above pseudo code we can see that the MPI communications have already been partially optimised, as they don't send a message for each element of the \verb|n| and \verb|ipde| loops, 
they aggregate the data to be sent or received into an array and then send that array, as shown in the following code (which implements one of the send steps in the pseudo code above):

\begin{verbatim}
  datasize = npde*((2*nharms)+1)
  tempindex = 1
  do n = 0, 2*nharms
    do ipde = 1, npde
      sendarray(tempindex,localsendnum) = 
&                 q2(in1,jn1,ipde,n)
      tempindex = tempindex + 1
    end do
  end do
  call sendblockdata(sendarray(1,localsendnum),iblk1,
&       iblk2,datasize,sendrequests(localsendnum))
\end{verbatim}

However, as the send and receive functionality is within a loop, and for that loop the send and receive processes do not change (the same sender and receiver are involved in all the communications 
for a given invocation of the loop) it is possible to reduce all these send and receives down to one send and one receive by further aggregating the data into a single send array 
and using a single receive array.

We performed a similar optimisation for the collective communications used in the code, where the original collective functionality was of the following form:
\begin{verbatim}
do i=1,nbody
   temparray(1) = cl(n,i)
   temparray(2) = cd(n,i)
   datalength = 2
   if(functag.eq.3) then
      temparray(3) = cm(n,i)
      datalength = 3
   end if
   call realsumallreduce(temparray,datalength)
   cl(n,i) = temparray(1)
   cd(n,i) = temparray(2)
   if(functag.eq.3) then
     cm(n,i) = temparray(3)
   end if
end do
\end{verbatim}

As well as a collective operation being undertaken for each iterations of the \verb|nbody| loop this functionality is also called from within a loop over harmonics (which sets the variable \verb|n| in the 
above code), enabling us to rationalise the number of collective operations undertaken from $nbody*((2*nharms)+1)$ to $1$ for each Runge-Kutta step in the simulation by collecting all the data to be sent into 
a single buffer, performing one collective all reduce, and unpacking the results at the end of the communication, as shown next:
\begin{verbatim}
j = 1
do k = 0,2*nharms
   do i=1,nbody
      temparray(j) = cl(k,i)
      j = j +1
      temparray(j) = cd(k,i)
      j = j + 1
      temparray(j) = cm(k,i)
      j = j + 1
   end do
end do
call realsumallreduce(temparray,j-1)
j =  1
do k = 0,2*nharms
   do i=1,nbody
      cl(k,i) = temparray(j)
      j = j +1
      cd(k,i) = temparray(j)
      j = j + 1
      cm(k,i) = temparray(j)
      j = j + 1
   end do
end do
\end{verbatim}
As with the previous communication aggregation we have performed this at the expense of extra memory requirements for the routine, however these are not significantly large so do not adversely impact the 
overall memory footprint of the code, even for high \verb|nbody| and harmonic sizes.

Finally, it was also evident from profiling that the MPI I/O functionality in COSA was consuming large amounts of runtime, especially as larger simulations were undertaken.  COSA produces a number of different output files, but for optimisation there are two types of file that are important, as they are the largest and require the most time to write; the flowtec files and the restart file.  COSA produces a single restart file at the end of the simulation (or more frequently if requested by the user) which can be used to restart the simulation from the point the restart file was written.   It also produces one flowtec file per real harmonic at the end of the simulation. The flowtec files contain the solution in a format suitable for use with the commercial CFD postprocessor and flow visualisation software TECPLOT.

When large simulations are executed the output can be extremely large, with the restart file being many gigabytes (GB) in size and each flowtec file being close to a GB in size.  The current code does use parallel I/O functionality, calling MPI I/O routines to perform the output from all processes at once.  However, the I/O is performed, as shown in the example below, through individual writes of data elements to the file one at a time:

\begin{verbatim}
  call setupfile(fid(n),disp,MPI_INTEGER)
  call mpi_file_write(fid(n), 4*doublesize,1,
&   MPI_INTEGER,MPI_STATUS_IGNORE,ierr)
  disp = disp + integersize 
  call setupfile(fid(n),disp,MPI_DOUBLE_PRECISION)
  call mpi_file_write(fid(n),x(i,j,n),1,
&   MPI_DOUBLE_PRECISION,MPI_STATUS_IGNORE, ierr)
  disp = disp + doublesize
  call setupfile(fid(n),disp,MPI_DOUBLE_PRECISION)
  call mpi_file_write(fid(n),y(i,j,n),1,
&   MPI_DOUBLE_PRECISION, MPI_STATUS_IGNORE, ierr)
  disp = disp + doublesize
  call setupfile(fid(n),disp,MPI_DOUBLE_PRECISION)
  call mpi_file_write(fid(n),rho,1,
&   MPI_DOUBLE_PRECISION, MPI_STATUS_IGNORE, ierr)
  disp = disp + doublesize
  call setupfile(fid(n),disp,MPI_DOUBLE_PRECISION)
  call mpi_file_write(fid(n),ux,1,
&   MPI_DOUBLE_PRECISION, MPI_STATUS_IGNORE, ierr)
  disp = disp + doublesize
  call setupfile(fid(n),disp,MPI_INTEGER)
  call mpi_file_write(fid(n), 4*doublesize,1,
&   MPI_INTEGER,MPI_STATUS_IGNORE,ierr)
\end{verbatim}

Where the \verb|setupfile| subroutine invokes the \verb|MPI_FILE_SEEK| function.  This use of MPI I/O is not optimal; generally MPI I/O gives the best performance when large amounts of data are written in a single call to the file.  However, the way the data is structured in COSA, and the format of the output files, currently prohibits doing this.  

It is important to the developers and users of COSA that the output files of the serial and parallel version of the code are the same therefore in the scope of this work it was not possible to change the way it currently writes the data.  However, we optimised the current functionality, aggregating the data to be written into arrays and then writing that data all at once.  An example of this optimisation of the I/O code outlined above is provided below:
\begin{verbatim}
  call setupfile(fid(n),disp,MPI_INTEGER)
  call mpi_file_write(fid(n), 4*doublesize,1,
&   MPI_INTEGER,MPI_STATUS_IGNORE,ierr)
  disp = disp + integersize 
  tempdata(tempindex) = x(i,j,n)
  tempindex = tempindex + 1
  tempdata(tempindex) = y(i,j,n)
  tempindex = tempindex + 1
  tempdata(tempindex) = rho
  tempindex = tempindex + 1
  tempdata(tempindex) = ux
  tempindex = tempindex + 1
  call setupfile(fid(n),disp,MPI_DOUBLE_PRECISION)
  call mpi_file_write(fid(n),tempdata(1),tempindex-1,
&   MPI_DOUBLE_PRECISION, MPI_STATUS_IGNORE, ierr)
  disp = disp + doublesize*(tempindex-1)
  call setupfile(fid(n),disp,MPI_INTEGER)
  call mpi_file_write(fid(n), 4*(tempindex-1),1,
&   MPI_INTEGER,MPI_STATUS_IGNORE,ierr)
\end{verbatim}

\subsection{Hybrid Optimisations}
Using profiling information from COSA we worked to optimise the OpenMP functionality by focussing on the core 
routines that are heavily used for these types of simulations.  This necessitated removing OpenMP functionality from areas of code that were not heavily used, and therefore removing the 
OpenMP overheads for those sections of the code,  and re-implementing some of 
the existing OpenMP functionality, including replacing some heavily used small shared arrays with private variables and reduction operations. 

Furthermore, the loops in the code that compute over the harmonics of the simulation are contained within subroutines which are called for each block in the simulation, and in each subroutine there 
can be a number of separate loops over harmonics.  Simply parallelising each harmonic loop with an OpenMP parallel do directive meant that there were a lot of places that OpenMP parallel regions
 are started and finished in the code.  There is an overhead associated with starting and finishing a parallel region in OpenMP, therefore we re-engineered the OpenMP code we had added to reduce these overheads 
by hoisted OpenMP parallel regions to higher levels in the program (where appropriate).

We also implemented {\it first touch} initialisation functionality to ensure that data is initialised on the cores that will be processing it.  The original code zeros all the data arrays when they are
allocated, and the allocation is done by the MPI process.  We have altered the zeroing of the arrays so it is done in parallel, following the parallelisation pattern that is used in the rest of the code.

Finally, there were also a number of parts of the code that had not been parallelised with OpenMP, primarily the I/O and MPI message passing code.  The MPI communications are performed over a loop of the cut, or halo, data.
Each cut is independent so they can be performed by separate threads.  However, as they involve MPI communications then we need to ensure that we are using the threaded version of the MPI library using the function 
\verb|MPI_INIT_THREAD| rather than the usual \verb|MPI_INIT| function.  Furthermore, we need to ensure that the MPI library being used can support individual OpenMP threads performing MPI 
communications (\verb|MPI_THREAD_MULTIPLE|).

The MPI communication code within COSA also implicitly depends on the order that the messages are send and received to match data with its correct location on the receiving process.  Each cut is processed in order 
and the individual sections of the cut are sent sequentially.  Therefore, simply parallelising this process using OpenMP will not guarantee that data is placed in the correct arrays once it is received.  To 
address this issue we added extra functionality to calculate where the data for each part of the cut should be placed in the send and receive arrays for each processes communications, and combined this with 
a unique identifier in the MPI tag for each message, to enable threads to send and receive the MPI messages as required and ensure that they place the received data in the correct places in the data arrays.

The I/O undertaken through the MPI code used MPI I/O functionality.  In general the I/O operations are independent for each block and each harmonic within the block.  However, there are a number of collective 
operations (operations that all processes must be involved in) in the I/O functionality, particularly opening and closing files.  To enable the OpenMP threads to be able to write to the restart and flowtec 
files independently we needed to ensure that all the threads are involved in the opening of the files so they each have a separate file handle to write.  Therefore, we implemented a hybrid file opening and 
closing strategy as follows:
\begin{verbatim}
!$OMP DO ORDERED
     do i=1,omp_get_num_threads()
!$OMP ORDERED
        call openfile(fid,'restart',iomode)
!$OMP END ORDERED
     end do
!$OMP END DO
\end{verbatim}

Where \verb|openfile| calls \verb|MPI_FILE_OPEN|.  The only other modification that need to be 
made to enable file writing from the OpenMP threads was to ensure that they could correctly calculate where the data for each harmonic need to be written to (rather than each block as was the case previously).

\section{Experimental Setup}
\label{sec:5}

We evaluated the performance of the MPI and hybrid parallelisations of COSA, and the optimisations we have undertaken, using a range of common 
large scale HPC platforms and a range of representative test cases.

\subsection{Test cases}

Two different simulations were used to evaluate the performance of COSA, outlined in the following sections.

\subsubsection{Test case 1}

This first test case is a HB analysis of a heaving and pitching wing designed to extract energy from an oncoming air stream. 
The 512-block grid has 262,144 cells, and 31 real harmonics are used. This HB analysis has the same memory requirements 
of a steady flow analysis with more than 8 million cells.  Further details on the aerodynamics of this device and the 
analysis of its efficiency based on COSA time-domain simulations are reported in the articles [1,3].

\subsubsection{Test case 2}

The other test case is a HB flow analysis of the blade section at 90\% span of a multi-megawatt horizontal axis wind turbine 
operating in yawed wind. The analysis has been performed using a fine grid with 2048 blocks, and 4,194,304 cells. 
In the simulations we have used 17 real harmonics. Further details on the time-domain and HB COSA analyses of this problem are reported in[2].

\subsection{Computing Resources}

We used three different large scale HPC machines to benchmark performance.  The first was a {\bf Cray XE 6}, HECToR, which is the UK National Supercomputing Service consists of 2816 nodes, each containing two 
16-core 2.3 GHz {\it Interlagos} AMD Opteron processors per node, giving a total of 32 cores per node, with 1 GB of memory per core. 
This configuration provides a machine with 90,112 cores in total, 90TB of main memory, and a peak performance of over 800 TFlop/s.  We used the PGI FORTRAN compile on HECToR, compiled with the \verb|-fastsse| 
optimisation flag.

The seconds was a {\bf Bullx B510}, called HELIOS, that is based on Intel Xeon processors.  A node contains 2 Intel Xeon E5-2680 2.7 GHz processors giving 16-cores and 64 GB memory.  HELIOS 
is composed of 4410 nodes, providing a total of 70,560 cores and a peak performance of over 1.2 PFlop/s.  The network is built using Infiniband QDR non-blocking 
technology and is arranged using a fat-tree topology.  We used the Intel FORTRAN compiler on HELIOS, compiling with the \verb|-O2| optimisation flag.

The final resource was a {\bf BlueGene/Q}, JUQUEEN at Forsch\-ungszentrum Juelich, which is and IBM BlueGene/Q system based on the IBM POWER architecture.
There are 28 racks composed of 28,672 nodes giving a total of 458,752 compute cores and a peak performance of 5.9 PFlop/s. 
Each node has an IBM PowerPC A2 processor running at 1.6 GHz and containing 16 SMT cores, each capable of running 4 threads, 
and 16 GB of SDRAM-DDR3 memory.   IBM's FORTRAN compile, xlf90, was used on JUQUEEN, compiling using the \verb|-qhot -O3 -qarch=qp -qtune=qp| optimisation flags.

Further technical details of these systems are provided in Table \ref{tab:hardware}.  The power listed in the table is the nominal power per node 
calculated using the reported power consumed during the LINPACK benchmark for the Top500\footnote{http://www.top500.org} list entries divided by the number of 
nodes in the system.
\begin{savenotes}
\begin{table}[th]
\small{
\centering
\caption{Details of the computer hardware used for the benchmarking}
\begin{tabular}{|p{100pt}|p{50pt}|p{30pt}|p{30pt}|} \hline
 & BlueGene/Q & Cray XE6 & Bullx B510 \\ \hline \hline
Processor & IBM PowerPC A2 & AMD Opteron 6276 & Intel Xeon E5-2680 \\ \hline
Processor Frequency (GHz) & 1.6 & 2.3 & 2.7 \\ \hline
Processor Cores & 16 & 16 & 8 \\ \hline
Processors per Node & 1 & 2 & 2 \\ \hline
Number of Cores & 458,752 & 90,112 & 70,560 \\ \hline
Memory per node (GB) & 16 & 32 & 64 \\ \hline
Peak performance per node (GFlop/s) & 204.8 &  294.4 &  345.6 \\ \hline
Power (Watt) & 80 & 400\footnote{The HECToR Top500 entry does not include power data so the Cray XE6 power figure is calculated using the data from the Gaea C2 entry in the list 
which is a comparable Cray XE6} & 498 \\ \hline
Network latency\footnote{
The network latency and bandwidth figures have been calculated used the mpbench benchmark 
which is part of the llcbench benchmark suite\cite{llcbench}} ($\mu$s) & 1.4 & 1.2 & 0.6 \\ \hline
Network bandwidth per node (GB/s) & 3.4 & 5.6 & 3.0 \\ \hline 

\end{tabular}
}
\label{tab:hardware}
\end{table}
\end{savenotes}

\section{Performance Results}
\label{sec:6}

We evaluated the effect of our performance optimisation on COSA on the different systems using a range of
MPI process counts per shared memory node on each system, and also comparing the best performing MPI benchmarks 
with the hybrid code using four OpenMP threads for every MPI process used.  We also used different iteration counts 
for the test cases to enable evaluation of the performance across a full range of node counts (for the small number of iterations) 
and for a more realistic usage scenario on a reduced number of nodes (using a large number of iterations).  Ideally the code 
would have benchmarked using a large number of iterations for all node counts, however the restrictions on the job queues on the 
systems we were using meant that the longest job we could run was 12 hours which limited the number of iterations that could 
be completed within this limit on a single node for the slowest system used.

There are two performance metrics we are evaluating for this code, overall time to solution and cost.  The user is generally 
looking to obtain results as quickly as possible, but may also be interested in getting results as efficiently as possible so they can 
utilise the HPC resource they have in the more efficient way.  Because each of the systems we are using for the benchmarking 
uses different processor, memory, network, and disk technology it is difficult to directly compare the performance based purely 
on runtime between the systems.  Therefore, we are evaluating relative performance (and therefore cost) using the estimated electrical power required for an iteration of the simulations on each system.

\subsection{MPI Results}

\begin{figure*}[t]
\centering
\subfloat[MPI Parallelisation]{\epsfig{file=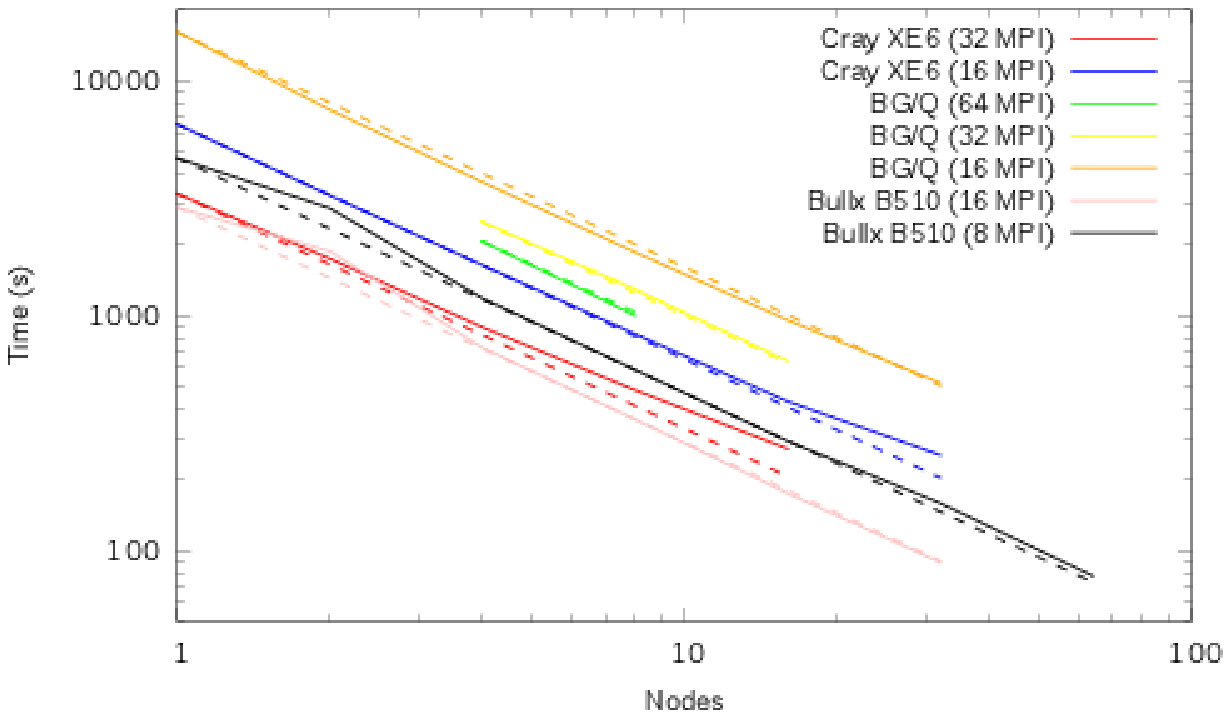,width=0.5\textwidth}\label{fig:runtime100iter}}
\subfloat[Hybrid Parallelisation]{\epsfig{file=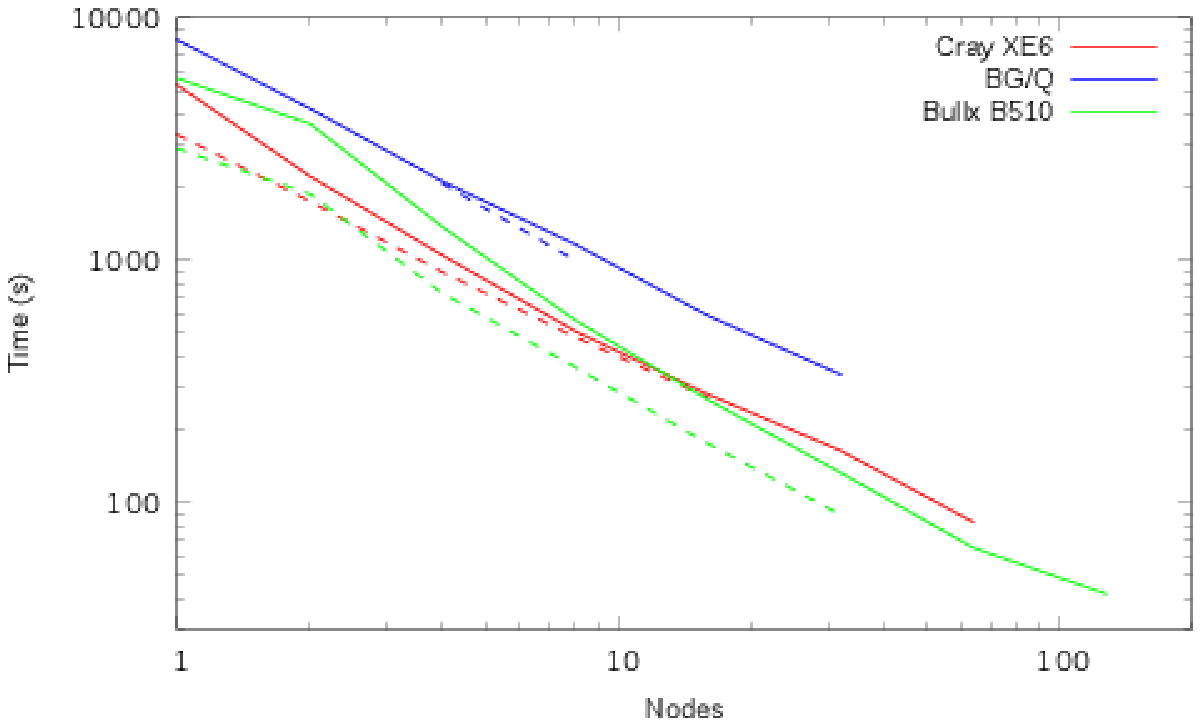,width=0.5\textwidth}\label{fig:hybridruntime100iter}}
\caption{Runtime of test case 1 using 100 iterations}
\end{figure*}

Figure \ref{fig:runtime100iter} presents the runtime of the optimised MPI code on all three systems, using 100 iteration of test case 1, for fully populated 
and under populated nodes.  The Cray XE6 has 32 cores per node, meaning a full populated node has 32 MPI processes per node.  The Bullx B510 
has 16 MPI processes per node when fully populated.  The BG/Q is slightly different as it has 16 hardware cores per node, but each core can run 4 separate threads very efficiently (which means each core can run 4 MPI processes), therefore a fully populated BG/Q node can have either 64 or 16 MPI processes depending on how a user wishes to utilise the hardware.
The dotted lines on the graph represent the ideal runtimes for their corresponding case based on the single node runtime figure.

We can compare the performance of the different machines using the graph in Figure \ref{fig:runtime100iter}.  We can see that, in terms of 
time to solution, the Bullx B510 gives the best performance.  It is significantly faster than the Cray XE6, even when using the same number of nodes 
(for instance 16 nodes on each system) which involves using double the number of cores on the Cray compared to the Bull machine.  We can also see 
that underpopulating nodes on the Cray and Bull does not improve performance, with underpopulated nodes requiring twice the number of nodes to 
achieve the same performance as the fully populated case.

The scaling of the code is better on the Bull (with the exception of the transition from 1 to 2 nodes) than on the Cray, and the scaling is also good 
on a node basis on the BG/Q system.  Underpopulating nodes on the BG/Q does significantly improve performance, reducing the time to solution for the 
simulation at the cost of extra resources.  However, we can see that when the BG/Q nodes are fully populated (i.e. using 64 MPI processes per node) 
we get better resource usage compared to the underpopulated case.  If we compare the 64 MPI processes per node case with the 16 MPI processes per node 
results we can see that the same time to solution is achieved using 8 nodes (when fully populating) compared to 16 nodes (when underpopulating by a factor of 4).
The downside of fully populating BG/Q nodes is that it is not always possible to run a given simulation on a set of fully populated nodes.  For instance, 
it was only possible to run test case 1 on 4 or 8 nodes, few nodes and there was not enough memory to execute the simulations.  The memory effect experienced is likely due the memory consumption associated with the MPI library rather than the COSA application itself.

\begin{figure*}[t]
\centering
\subfloat[MPI Parallelisation]{\epsfig{file=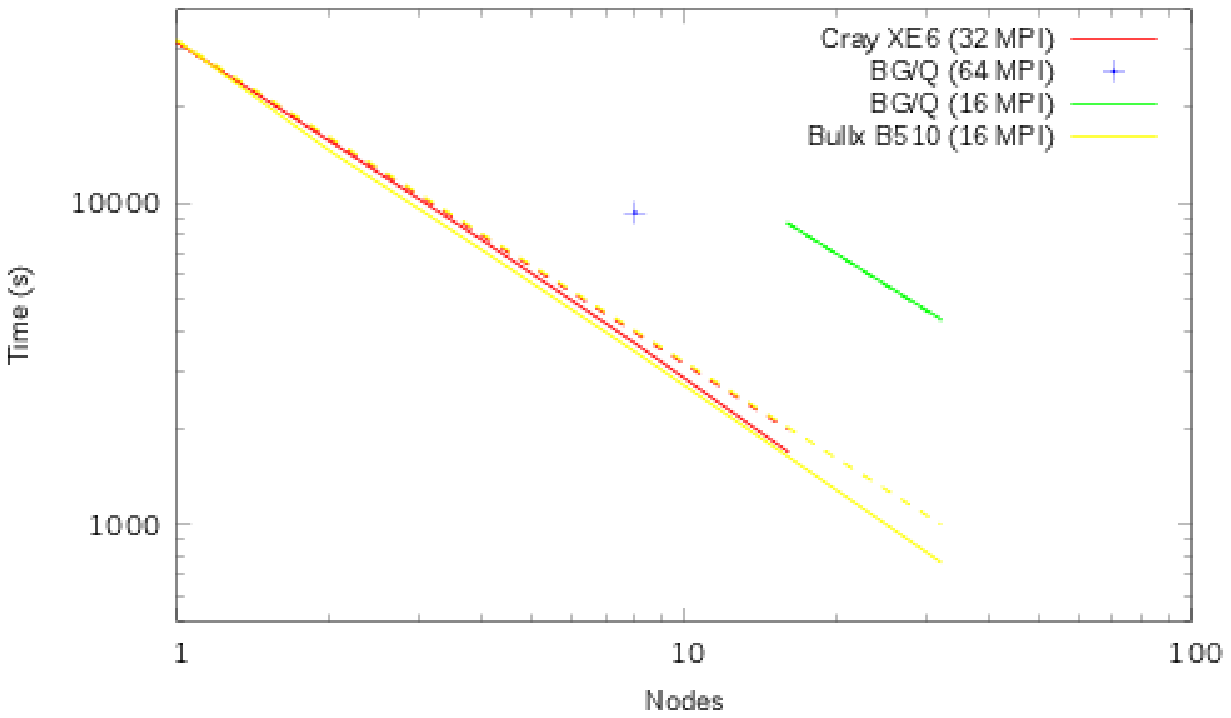,width=0.5\textwidth}\label{fig:runtime1000iter}}
\subfloat[Hybrid Parallelisation]{\epsfig{file=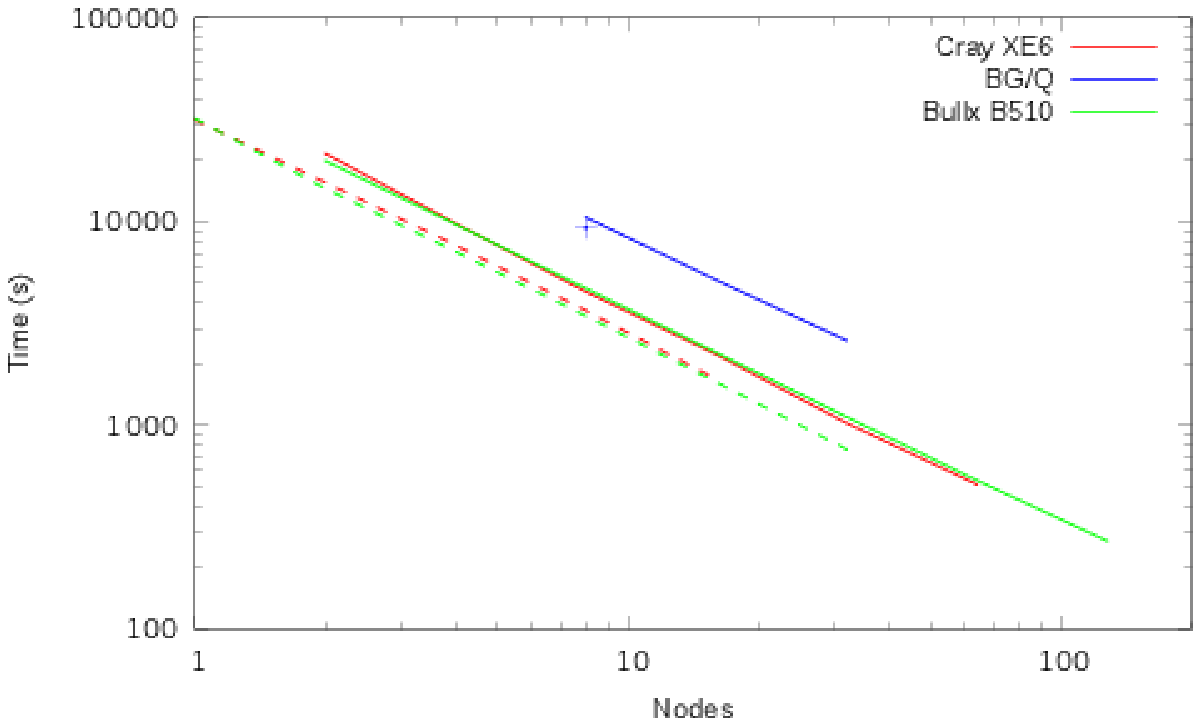,width=0.5\textwidth}\label{fig:hybridruntime1000iter}}
\caption{Runtime of test case 1 using 1000 iterations}
\end{figure*}

We also ran the same simulation with a large number of iterations to evaluate the efficiency of the main computational part of the code (increasing the iterations 
reduces the impact of the initial and final I/O functionality on the overall runtime).  Figure \ref{fig:runtime1000iter} presents the time to solution for the 
same testcase with 1000 iterations.  Now we can see the overall runtime scaling has improved, with all three 
architectures exhibiting better than ideal scaling (on a node level scaling).  For 1000 iterations it was 
only possible to run on the BG/Q with 64 processes per node on the maximum number of nodes for that test cases (8 nodes) as any fewer nodes required longer runtime than the queuing system would allow on that particular machine.  However, we can see that the 8 node runtime for fully populated BG/Q has approximately the same runtime as 16 nodes of under-populated BG/Q.

\begin{table}[t]
\centering
\caption{Node scaling efficiency of the parallelisations}
\begin{tabular}{|l|c|c|c|} \hline
 Architecture & Code & Iterations & Efficiency \\ \hline \hline
BlueGene/Q & MPI & 100 & 103\% \\
  & Hybrid & 100 & 74\% \\
  & Hybrid & 1000 & 89\% \\ \hline
Cray XE6 & MPI & 100 & 76\% \\ 
 & Hybrid & 100 & 76\% \\
 & MPI & 1000 & 117\% \\ 
 & Hybrid & 1000 & 82\% \\ \hline
Bullx B510 & MPI & 100 & 101\% \\ 
 & Hybrid & 100 & 50\% \\
 & MPI & 1000 &  131\% \\ 
 & Hybrid & 1000 & 70\% \\ \hline
\end{tabular}
\label{tab:efficiency}
\end{table}

We also present the efficiency of the node scaling of the code on the different machines in Table \ref{tab:efficiency}.  The efficiency of the MPI code is calculated using the following 
equation:

\begin{equation}
  EM_n = \frac{\frac{TM_s}{TM_n}}{\frac{n}{s}}
  \label{eq:mpiefficiency}
\end{equation}

Where $EM_n$ is the efficiency of the MPI code at $n$ nodes, $TM_s$ is the runtime on $s$ (the smallest number of nodes used for the benchmark), and $TM_n$ is the runtime on $n$ nodes.
Note that on the BG/Q we could only run the 1000 iteration version of test case 1 on 8 nodes so we cannot calculate the scaling of this test case.  We are using the 
most efficient configurations of these systems for this table (64 MPI processes per node on BG/Q, 32 MPI processes per node on Cray XE6, 16 MPI processes per node on Bullx B510).

We can see that, especially when using a larger number of iterations, we achieve excellent node scaling efficiency with the optimised COSA MPI code.  

\subsection{Hybrid Results}

The hybrid code was run using the same test cases and the performance compared with the MPI code.  Figure \ref{fig:hybridruntime100iter} presents the runtime for the 
hybrid code using a range of nodes.  The dotted lines on this plot are the best MPI runtime from the respective systems (64 MPI processes per node for the BG/Q; 32 and 16 
processes per node for the Cray and Bull machines respectively).  It is evident from this graph that the hybrid code can provide improved time to solution over the MPI code 
for all the architectures.  On the BG/Q the hybrid code exhibits similar performance to the MPI code for the comparable node counts, and reduces the runtime by around 3 times when 
4 times the number of nodes is used.  For the Cray and Bull machines the hybrid code has significantly lower performance than the MPI code for comparable node counts, but still 
reduces time to solution compared to the MPI code when more resources are used (3.26 times faster for the Cray and 2.11 times fast for the Bull when using 4 times the resources).

As with the MPI benchmarking we have evaluated the performance using 1000 iterations instead of 100 iterations, as shown in Figure \ref{fig:hybridruntime1000iter}.  As with the MPI evaluation, increasing the iterations has improved the overall scaling of the hybrid code, bring the results closer to those of the pure MPI code (shown as the dotted lines the graph). 

We have also evaluated the efficiency of the Hybrid code is calculated using the following equation:

\begin{equation}
  EH_n = \frac{\frac{TM_n}{TH_m}}{\frac{m}{n}}
  \label{eq:hybridefficiency}
\end{equation}

Where $EH_n$ is the efficiency of the hybrid code at $m$ nodes, $TM_n$ is the runtime of the best MPI code on $n$ (the maximum nodes it can use for test case 1), and $TH_m$ is the runtime of the 
hybrid code on $m$ nodes (the maximum number of nodes the hybrid code can exploit using 4 OpenMP threads per MPI process using test case 1).

As shown in Table \ref{tab:efficiency} we can see the hybrid code does not show as good 
performance as the MPI code, but is still giving acceptable performance for the Cray and BG/Q architectures considering that we are using four times the resources as the MPI code, and therefore 
we are reducing the amount of work for each thread to undertake by a factor of 4.

\subsection{Resource Efficiency}

As previously discussed, to enable us to evaluate the performance of COSA across a range of resources we are looking at the notional power consumption of the code 
when running a simulation.  To perform this analysis we are using the power figures presented in Table \ref{tab:hardware}, which as previously explained is the 
amount of power reported consumed by each node during LINPACK benchmarking.  We have used this figure, along with the runtime of the simulation, to calculate the 
power consumed per iteration of the simulation, as shown in equation \ref{eq:power}:

\begin{equation}
  P_i = \frac{\frac{T_t}{3600} * P_n}{ni}
  \label{eq:power}
\end{equation}

Where $P_i$ is the power consumed per iteration in Watt hours, $T_t$ is the total runtime of the simulation in seconds, $P_n$ is the notional power per node figure, and $ni$ is the number of iterations undertaken.

\begin{figure*}[t]
\centering
\subfloat[MPI Parallelisation]{\epsfig{file=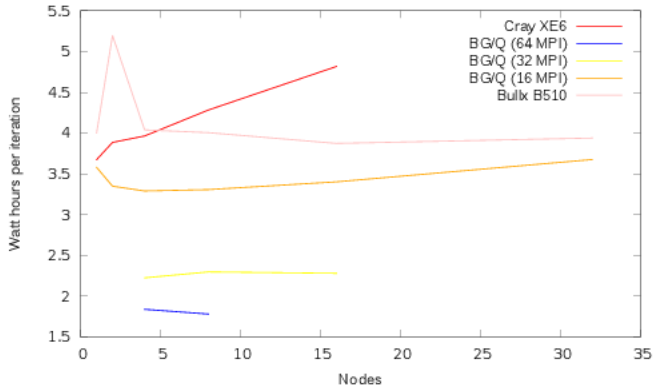,width=0.5\textwidth}\label{fig:power100iter}}
\subfloat[Hybrid Parallelisation]{\epsfig{file=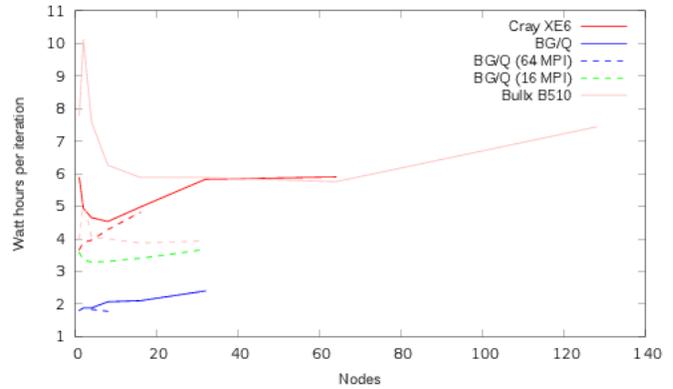,width=0.5\textwidth}\label{fig:hybridpower100iter}}
\caption{Power usage of test case 1 using 100 iterations}
\end{figure*}

Figure \ref{fig:power100iter} displays the relative power usage of the simulation on the three systems.  We can see that the BG/Q has the best power performance for the simulations we have undertaken, 
requiring less power to complete the simulation for all the configurations of fully populated or underpopulated nodes when compared to the other two systems.  It is also evident that fully populating the 
nodes makes much more efficient use of the resources, albeit for an increased time to solutions.

We should note that these are estimated power figures, not recorded power usage, so the actual power consumed may vary when under populating nodes, which may mean the figures for the BG/Q 16 and 32 MPI processes 
per node are not accurate.

The Bull machine shows generally fixed power efficiency regardless of the number of nodes used (disregarding the 2 node results), which highlights the MPI code is scaling well across nodes on this machine, whereas 
the power consumed by the Cray increases as we increase the number of nodes, highlighting sub-optimal scaling of the code in this instance.  We can also observe that the Cray requires less power than the Bull machine 
for small numbers of nodes, and indeed is close to the performance of the underpopulated BG/Q using only one node (even though the Cray completes the 1 node run nearly 5 times faster than the BG/Q), but as we scale 
the number of nodes the efficiency of the code on the Cray gets progressively worse, becoming less efficient than the Bull machine when going from 4 to 8 nodes.

When considering the power efficiency of the hybrid code, shown in Figure \ref{fig:hybridpower100iter} (where the dotted lines are the power per iteration of the most efficient MPI runs for comparison), we can see that for the Cray and 
Bull machines the hybrid code generally has considerably lower efficiency than the pure MPI code, with the hybrid code on the Cray now showing similar or better efficiency when compared to the hybrid code on the Bull, although 
the hybrid code on the Cray has similar efficiency compared to the pure MPI code when using 8 and 16 nodes, whereas on the Bull machine the efficiency is always much worse than the pure MPI code.

On the BG/Q the picture is different.  Here the efficiency of the hybrid code still does not beat that of the pure MPI code, however it is much closer, and if compared to the underpopulated BG/Q MPI result 
(where 16 MPI processes are used per node) it is considerably better.  Given that the hybrid code on BG/Q enables a user to exploit as many nodes as the underpopulated approach takes (4 times the number of nodes that 
the fully populated MPI code can use for this code) at an efficiency that is close to that of the fully populated case we can see that the hybrid code is extremely beneficial on the BG/Q.

\begin{figure*}[t]
\centering
\subfloat[MPI Parallelisation]{\epsfig{file=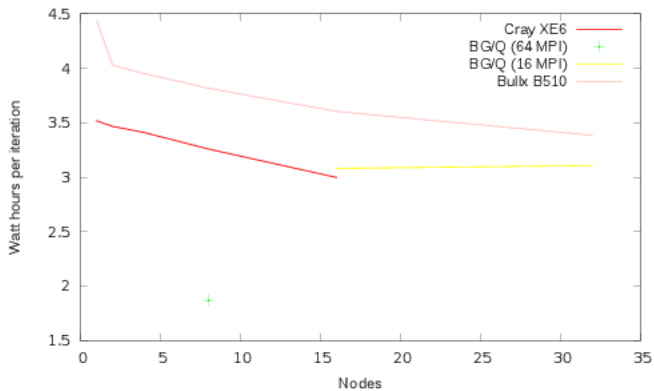,width=0.5\textwidth}\label{fig:power1000iter}}
\subfloat[Hybrid Parallelisation]{\epsfig{file=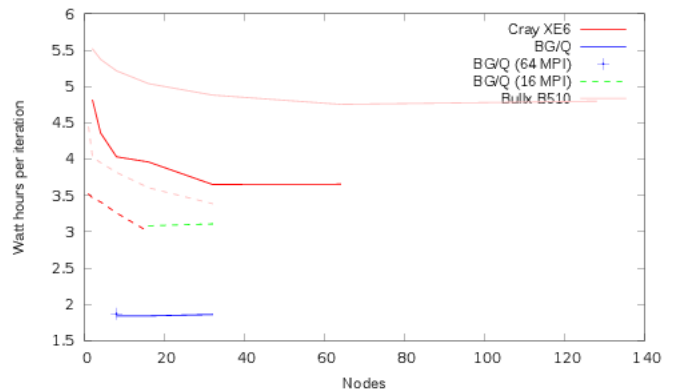,width=0.5\textwidth}\label{fig:hybridpower1000iter}}
\caption{Power usage of test case 1 using 1000 iterations}
\end{figure*}

If we consider the efficiency of the code using 1000 iterations for the same test case we can see a further improved picture, as we so with the runtime scaling of the code in the previous section.
We can now see that the efficiency of the Cray has much improved, no longer increasing with as the nodes increase.  A similar effect is also exhibited by the Bull and IBM machines, and it is also evident that the increased iterations have reduced the gap between the power consumption of the hybrid implementation on the Bull and Cray with that of the MPI implementation.  We can also see that the performance of the hybrid code on the BG/Q is also much improved.  The hybrid implementation uses approximately the same power as the fully populated MPI code despite scaling out to 4 times the node and reducing the runtime by around 3.5 times.

\subsection{Large Benchmark}

We have also evaluated the optimisations using test case 2 on the different computing resources we have access to.  Test case 2 can utilise 2048 MPI processes using the pure MPI code, but only has half the real harmonics that test case 1 had so we are only utilising 2 OpenMP threads for the hybrid case, meaning we can use up to 4096 cores for the hybrid case.   When we evaluate the optimised codes using this test case we get a scaling efficiency of around 80\% efficiency on the Cray and more than 90\% efficiency on the Bull (we could only run one test on the BG/Q as lower node counts would not run for the fully populated configuration).  The BG/Q required between 4 and 5 times less energy per iteration than the other two machines.

The hybrid version of this code exhibited around 80\% efficiency when using 2 OpenMP threads per process for the machines used in this evaluation.

\section{Conclusions}

Through the optimisation we have undertaken on the parallelisations of COSA we are able to ensure that the code scales with excellent efficiency as the number of nodes are increased for the MPI parallelisation, 
with around or exceeding 100\% efficiency across a range of systems providing a realistic number iterations are used in the simulation.  It should be noted that this performance is including the full functionality 
of the code, including input and output of data.

We have demonstrated that on hardware which is designed for explicit multi-threading, such as the BG/Q, the hybrid code gives excellent performance when scaling beyond the number of nodes that the MPI code 
can used, enabling users to efficiently reduce the time to solution at almost ideal efficiency.  When comparing to more traditional hardware, such as used in the AMD based Cray or the Intel based Bull machines 
the hybrid code does not provide as good efficiency when scaling the number of nodes.  However, it should be noted that the hybrid code is, but it's nature, going to be most efficient when using more resources 
than the pure MPI code.  This is because the hybrid code undertakes parallelisation of the harmonics of each block in the simulation.  If an MPI process has more than one block then it will encounter OpenMP overheads 
for each block it processes, whereas if the MPI code is maximally parallelised (i.e. 1 block per MPI process) then these overheads are minimised.  If we examine the performance of the hybrid code in that scenario 
we can see that we are achieving between 70\% and 90\% efficiency, enabling time to solution to be significantly reduce for a minimal reduction in efficiency.

We have also observed the the BG/Q architecture gives significantly better power to iteration performance than the other two systems, albeit for a longer overall time to solution.  However, the low memory per thread 
on the BG/Q (256 MB per thread when using 64 threads per node) is a significant barrier to memory applications utilising such a system, but we have shown that a hybrid approach can alleviate this problem by 
enabling all the resources on the node to be used by a code without having to have 64 MPI processes per node (with all the associated fixed memory requirements those MPI processes have).

We have further work to undertake to understand the performance difference between the Cray and Bull systems, particularly why the Bull system exhibits poorer hybrid performance than the rest.  However, a 
preliminary hypothesis is that the higher performance of the individual nodes means that the test cases we are using are not sufficiently computationally demanding on the Bull system to warrant the hybrid parallelisations.
It is also possible that the ratio of MPI processes to OpenMP threads was not optimal for this system.  Furthermore, we have not investigated the thread assignment behaviour of any of these systems, simply 
using what is provided by default through the batch systems (assigning the requisite MPI processes per node and OpenMP threads per process) so it is possible that non-optimal thread and process binding in affecting 
performance.

\section{Acknowledgments}
This work was supported by Dr M. Sergio Campobasso at Lancaster University.

Part of this work was funded under the HECToR Distributed Computational Science and
Engineering (CSE) Service operated by NAG Ltd. HECToR -- A Research Councils
UK High End Computing Service -- is the UK's national supercomputing service,
managed by EPSRC on behalf of the participating Research Councils. Its mission is 
to support capability science and engineering in UK academia. The HECToR
supercomputers are managed by UoE HPCx Ltd and the CSE Support Service is
provided by NAG Ltd. http://www.hector.ac.uk

Part of this work was supported by NAIS(the Centre for Numerical Algorithms and Intelligent Software) which 
is funded by EPSRC grant EP/G036136/1 and the Scottish Funding Council.

\balancecolumns
\end{document}